\documentclass[aps,pra,twocolumn,showpacs]{revtex4}

\usepackage{epsfig}

\begin{document}


\title{Semiclassical Theory of Coherence and Decoherence}

\author{Gregory A. Fiete$^{1,2}$ and Eric J. Heller$^{1,3}$}
\affiliation{$^1$Department of Physics, Harvard University, Cambridge, 
        MA 02138\\$^2$Kavli Institute for Theoretical Physics,
University of California, Santa Barbara, CA 93106\\$^2$Department of
Chemistry and Chemical Biology, Harvard University, Cambridge, MA
02138}


\begin{abstract}
A general semiclassical approach to quantum systems with system-bath
interactions is developed.  We study system decoherence in detail
using a coherent state semiclassical wavepacket method which avoids
singularity issues arising in the usual Green's function approach.  We
discuss the general conditions under which it is approximately correct
to discuss quantum decoherence in terms of a ``dephasing'' picture and
we derive semiclassical expressions for the phase and phase
distribution.  Remarkably, an effective system wavefunction emerges
whose norm measures the decoherence and is
equivalent to a density matrix formulation. 
\end{abstract}
\maketitle

\section{Introduction}
\label{sec:intro}
The challenge of understanding to what extent a quantum system can
retain its coherence in the presence of interactions with other
degrees of freedom has attracted much attention.  Much of it is
motivated by recent advances in mesoscopic\cite{ji,sohn, imry:book}
and cold-atom experiments\cite{inguscio, ketterle} as well as keen
interest in quantum computing, which depends crucially on quantum
coherence\cite{divincenzo, nielson}.

An essential ingredient in any discussion of coherence and decoherence
is the identification of a ``system'' and a ``bath'' which interact
with each other in such a way that a meaningful distinction can be
made between the two.  In the double slit experiment, for example, the
electrons (or other quantum objects, e.g., ``bucky balls''\cite{nairz}) are
taken as the system and the degrees of freedom in the slits (phonons
or spins, for example) are taken as the bath.  Because the experiment
only involves detecting the interference pattern on a screen behind
the slits, no direct measurement of the bath (i.e. the slit degrees of
freedom) is made. (Since one has no knowledge of the state of the bath
one must sum over all possible states of the bath, i.e. one ``traces
over the bath''.) Only the system is directly observed.  As is well
known\cite{feyn2}, if the bath detects the path of the particle no
interference pattern will be seen; the particle has therefore
decohered.  On the other hand, if there is no or only partial
detection of the path of the particle by the bath, some interference
pattern will be seen with its intensity reflecting the degree of
coherence of the particle\cite{feyn2}.  We will later show how these
familiar statements appear in a very transparent way in our
semiclassical formalism.

The problem of a quantum system interacting with an environment has
been addressed many times in the literature, e.g.
\cite{feynman,caldeira,ady}.  The Feynman-Vernon influence functional
approach is well known, although its usefulness beyond the context of
harmonic baths has been an issue.  The influence functional approach
to more realistic systems has been advanced significantly by Makri and
Thompson\cite{makri1,makri2}, exploiting and developing coherent state
methods with smooth kernels suitable for Monte Carlo sampling.
However, the generality of their approach necessarily means some
detailed insights and limiting cases are lost in the machinery, so to
speak.

Another approach is to make a semiclassical approximation for the
system-bath evolution. Casting the system-bath interaction and the
decoherence problem in terms of semiclassical wavepacket dynamics
proves to be a useful and insightful exercise.  Since a semiclassical
approach is based on classical trajectories, one can present an
intuitive picture of the criteria for coherence and decoherence in a
coupled system-bath. One may imagine a more traditional van Vleck
semiclassical Green's function approach; however, this has the
difficulty that caustic infinities abound in the van Vleck prefactors
due ultimately to the failure of the stationary phase approximation in
the limit of small action changes.  For example, suppose we consider a
harmonic oscillator and represent the $n^{th}$ state, $|n\rangle$,
semiclassically so that $\langle x|n\rangle$ has singularities at the
classical turning points for energy $E_n$.  Suppose now we displace
$|n\rangle$ slightly in position; call this $|\tilde n\rangle$. The
semiclassical projections onto all of the original basis states
$\langle n|\tilde n\rangle$ are completely wrong for small
displacements.  All but one of the projections are incorrectly
predicted to be $0$ since their classical manifolds do not overlap,
whereas the overlap with the undisplaced original state is nearly
singular.  The same displacement of the $n^{th}$ harmonic oscillator
state, expanded in terms of localized Gaussians, is quite accurate; it
has no such difficulties. Slight displacements of system or bath
states is commonplace in the decoherence problem, so we avoid the
caustic difficulties by starting with a wavepacket description,
avoiding the singularities.  We have called this the ``oil on troubled
waters'' effect of using a wavepacket description\cite{postmodern}.

One of the insights which emerges from this approach relates to recent
discussions in the literature concerning the equivalence of a ``bath
overlap'' picture of decoherence with a ``system dephasing''
picture. (See Stern, Aharonov and Imry
(SAI)\cite{ady,ady:nato,imry:book} and Feynman and
Vernon\cite{feynman}.)  We find that there are three main processes
that contribute to decoherence: (i) Phase jitter (ii) Bath overlap
decay and (iii) Shifts in the trajectory of the system wavepacket.  We
present explicit formulas for each of these effects within our
semiclassical wavepacket description.

This paper is organized as follows.  In Sec.~\ref{sec:theory} we
develop the main ideas of our paper and derive the general expression
for the coherence of a quantum system coupled to a general bath
(described by a density matrix), Eq.~(\ref{central}).  This expression
can be recast in the form of a very intuitive effective system
wavefunction, Eq.~(\ref{eq:eff_wavefunction}), which makes transparent
the effects of system-bath interactions on the system (described by
the norm of the effective wavefunction). In Sec.~\ref{sec:cases} we
study several important cases of Eq.~(\ref{central}) and
Eq.~(\ref{eq:eff_wavefunction}) in which certain physical process
(phase jitter, etc.)  dominate the system decoherence.  In
Sec.~\ref{sec:conclusions} we summarize our main results and
conclusions.  Important supplementary material is presented in the
appendixes.  In Appendix~\ref{app:gauss} we detail how to compute the
equations of motion perturbatively for Guassian wavepacket dynamics
and derive expressions needed in the main text.  In
Appendix~\ref{app:imry} we sketch the arguments of Stern, Aharonov and
Imry\cite{ady} which equate ``bath overlap'' and ``dephasing'' in a
special case system-bath interaction.

\section{Semiclassical Theory of Decoherence}
\label{sec:theory}
We set out to construct a general formal context for decoherence, with
the goal of reaching a useful and intuitive physical picture.  The
most general formal structure for decoherence (e.g. influence
functionals for general anharmonic baths) would not involve
semiclassical approximations, and could claim formal exactness.
However such formulations must necessarily miss the mark on the issue
of ``useful and intuitive''.

\begin{figure}
\centering
\includegraphics[width=8cm]{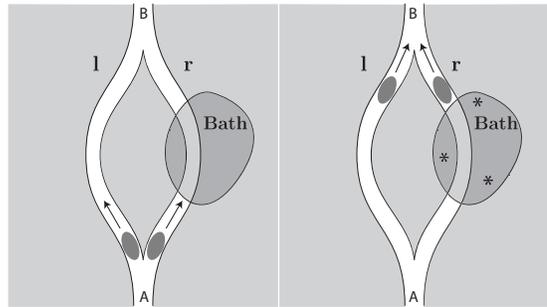}
\caption{A wavepacket representing the ``system'' in a narrow
waveguide is split coherently between two arms, one of which interacts
with a ``bath'' of particles. Upon recombining on the other side,
decoherence due to the system-bath interaction affects the
interference (or lack of it) between the two arms.}
\label{fig1}
\end{figure}

It is helpful to have a specific model in mind. The model of a two
armed device already introduced and used, for example, in the work of
Stern, Ahronov and Imry\cite{ady} serves that purpose well.  In
Fig.~\ref{fig1}, a wavepacket representing the system is coherently
split into two pieces, one of which later interacts with a bath.  The
degree of coherence can be checked experimentally by combining the
packets (as in the Aharonov-Bohm experiments of Ref.\cite{yacoby}),
although it is more general to check ${\cal M} \equiv {\rm Tr}[\hat
\rho_{\rm red}^2]$, where $\hat \rho_{\rm red}$ is the reduced density
matrix for the system, after tracing over the bath variables.

An ambiguity in checking for interference fringes of the recombined
beams is illustrated by supposing that there is a static potential
maximum in the left arm but not the right, and no system-bath
interaction in either arm.  This will cause a time delay of the left
wavepacket compared to the right, so no interference will result, even
though the system is completely coherent.  There are ways to avoid
this, such as taking longer coherent wave trains initially for the
system, but one must be careful, and other ambiguities can arise.  

On the other hand, ${\cal M}$ is simply $1$ for a completely coherent
system, and less than $1$ if some decoherence has taken place. For
example, suppose we have a system wavepacket broken into two non-overlapping
but coherent pieces, i.e.  $\psi = {1/\sqrt 2}(\psi^l + \psi^r)$, with
$\langle \psi^l\vert \psi^r\rangle = 0$ and $\langle \psi^l\vert
\psi^l\rangle = \langle \psi^r\vert \psi^r\rangle=1$.  Then the
density matrix, $\hat \rho$, is
\begin{equation} 
\hat \rho = {1\over 2} \sum\limits_{i,j} \vert \psi^i\rangle\langle \psi^j\vert \;,
\end{equation} 
with $i$ and $j$ taking on the values $l$ and $r$. It is easily seen
that $ {\rm Tr}[\hat \rho ]= {\rm Tr}[\hat \rho^2] = 1$, i.e. the
system is completely coherent.  However, if somehow the two parts $l$
and $r$ become completely decohered, (this in fact requires the action
of more degrees of freedom--a bath) we loose the off diagonal elements of
$\hat \rho$, getting
\begin{equation} 
\bar {\hat \rho} = {1\over 2} \left (\vert \psi^l\rangle\langle \psi^l\vert + \vert \psi^r\rangle\langle \psi^r\vert\right )\;.
\end{equation} 
Now we have $ {\rm Tr}[\bar {\hat \rho} ]= 1$, but $ {\rm Tr}[\bar
{\hat \rho}^2 ]= 1/2$.  

This is not the end of decoherence for this system, if the left and
right packets somehow undergo their own, ``internal'' decoherence.
This ``internal'' decoherence will happen on a much longer time scale
than the decoherence of the two initially coherent wavepackets because
it is the {\it difference} in the interactions that each wavepacket
experiences that determines the decoherence rate.  This rate is almost
always larger for two separated wavepackets than for a given
wavepacket.  All this is elementary, but it sets the stage for the
more detailed work to follow.

Turning now to a discussion of our general problem of a system
interacting with a bath, we cast our expressions in terms of density
matrices.  The most general density matrix for a bath expressed in
terms of Guassian wavepackets is
\begin{equation} 
\label{eq:d_matrix_bath}
\hat \rho_{\rm bath} = \sum_{i,j} w(i,j)\vert G_{i 0} \rangle\langle G_{j 0} \vert\;, 
\end{equation} 
where $w(i,j)$ satisfies all the properties necessary so that
$\hat \rho_{\rm bath}$ is a well-defined density matrix (${\rm
Tr}_{\rm bath} \left [\hat \rho_{\rm bath}\right ]=1,
\left[w(i,j)\right]^\dagger=w^*(i,j)=w(j,i)$, etc.).  In the special
case of a pure state bath, we can write $w(i,j)=w_i w_j
e^{i(\xi_i-\xi_j)}$ where the $w_i$ are real, positive numbers and the
$\xi_i$ are the associated phases.  
The states $\vert G_{i0} \rangle$
are multidimensional Gaussian wavepackets representing the bath
states.  Expanding the bath states in Gaussians allows us to make use
of very intuitive notions of classical mechanics (which guide the
wavepacket trajectories in the semiclassical approximation) while at
the same time permitting us to overcome technical difficulties
(singularities) in a more naive semiclassical treatment with van Vleck
propagators.

Assuming that we take our system be the wavepacket initially split into two
coherent pieces as shown in Fig.~\ref{fig1}, the initial wavefunction
of the system can be written as 
\begin{equation}
\vert \psi_{\rm sys}\rangle  ={1\over \sqrt 2}  
\left ( \vert g_0^l\rangle + \vert g_0^r \rangle \right )\;,
\label{eq:wp_sys} 
\end{equation}
where the states $\vert g_0^r\rangle$ and $\vert g_0^l\rangle$ are the
right (r) and left (l) Gaussian wavepackets (coherent states) near the
region A in Fig.~\ref{fig1}. The density matrix for the system is then
$\hat
\rho_{\rm sys}=\vert \psi_{\rm sys}\rangle \langle \psi_{\rm
sys}\vert$.  Expressed in terms of Eq.~(\ref{eq:wp_sys}),
\begin{equation}
\label{eq:d_matrix_sys}
\hat \rho_{\rm sys}= {1\over 2}\left( \vert g_0^l\rangle + \vert g_0^r \rangle \right ) \left(\langle g_0^l\vert + \langle g_0^r \vert \right )\;.
\end{equation}
The total initial density matrix of the system and bath is then 
\begin{equation}
\hat \rho_{\rm tot}(0)=\hat \rho_{\rm sys}\hat \rho_{\rm bath}\;.
\label{eq:rho_init}
\end{equation}
To study the decoherence of the system due to interactions with the
bath, we must compute the time evolution of Eq.~(\ref{eq:rho_init}).
Our approach is to use the the perturbative wavepacket time evolution
described in Appendix~\ref{app:gauss}.  The key result is that
the state
\begin{equation}
\left( \vert g_0^l\rangle + \vert g_0^r \rangle \right )\vert G_{i 0}\rangle \stackrel{time}{\longrightarrow} \vert g_{0 t}^l\rangle \vert G^l_{i t} \rangle + \vert g_{i t}^r \rangle \vert G^r_{i t}\rangle e^{i\phi_i}\;,
\label{eq:guassian_evolution}
\end{equation}
where $\vert g_{i t}^{r,(l)}\rangle$ is the wavepacket of the system
particle moving in the right (left) arm at time $t$ if the bath was in
state $\vert G_{i0} \rangle$ initially.  The state $ \vert
G^{r,(l)}_{i t} \rangle $ is the perturbed wavepacket of the bath
(according to Appendix~\ref{app:gauss}) at time $t$ for a particle
moving in the right (left) arm where the bath was initially in state
$\vert G_{i0} \rangle$.  Note that since the particle in the left arm
does not interact with the bath, $\vert G^l_{i t} \rangle=\vert
G^0_{it} \rangle$ and $\vert g_{i t}^{l}\rangle= \vert g_{0
t}^{l}\rangle$.  The phase $\phi_i$ is given by Eq.~(\ref{pha}).  

We emphasize that it is the local nature of the Gaussian wavepackets
combined with weak interactions that allows us to write down
Eq.~(\ref{eq:guassian_evolution}) with a {\it product state} for the
piece of the wavefunction that interacted with the bath on the right
arm.  This approximation actually becomes exact as $\hbar \rightarrow
0$. It is precisely the lack of ``local entanglement''
(Eq.~(\ref{eq:guassian_evolution}) still implies ``global
entanglement'' of course) in the wavefunction that makes our approach
conceptually convenient.  In a more general basis, we would have a sum
of terms for the right arm piece of the wavefunction and it would be
difficult to identify physical phases of the type given by
Eq.~(\ref{pha}).

The total density matrix at later times thus becomes
\begin{eqnarray}
\hat \rho_{\rm tot}(t)={1\over 2}\sum_{i,j} w(i,j)&&\left(\vert g_{0 t}^l\rangle \vert G^l_{i t}\rangle + \vert g_{i t}^r \rangle \vert G^r_{i t}\rangle e^{i\phi_i}\right)\nonumber \\
&&\times \left(\langle g_{0 t}^l\vert \langle G^l_{j t}\vert + \langle g_{j t}^r \vert \langle G^r_{j t} \vert e^{-i\phi_j}\right)\;,
\end{eqnarray}
which can be rewritten as
\begin{equation}
\hat \rho_{\rm tot}(t)=\hat \rho^{ll}(t) + \hat \rho^{lr}(t) +
\hat \rho^{rl}(t)+\hat \rho^{rr}(t)\;,
\end{equation}
where $\hat \rho^{ll}(t), \hat \rho^{lr}(t),$ etc. have the obvious
meaning. To study the coherence of the system, we trace over the bath
degrees of freedom to obtain the reduced density matrix,
  \begin{eqnarray} 
\hat \rho_{\rm red} &\equiv&  {\rm Tr}_{\rm bath} \left [\hat \rho_{\rm tot}  \right ] \nonumber \\
& =& 
 {\rm Tr}_{\rm bath} \left [\hat \rho^{rr}  
+ \hat \rho^{rl} + 
\hat  \rho^{lr} + \hat \rho^{ll} \right ]\;,\nonumber \\
\end{eqnarray}
which yields, for example, 
\begin{equation} 
\label{genred}
\hat \rho_{\rm red}^{rl} = {1\over  2} \sum\limits_{ij} \vert g_{i t}^r\rangle \langle g_{0 t}^l\vert w(i,j) e^{i \phi_i}{\cal O}_{ji}^{0r}\;,
\end{equation}
where 
\begin{equation} 
{\cal O}_{ji}^{0r} \equiv \langle G^0_{j t}\vert G_{i t}^r\rangle\;,
\end{equation} 
with a similar expression for $\hat \rho_{\rm red}^{lr}$ (just the
hermitian conjugate of $\hat \rho_{\rm red}^{rl}$) and the other
terms. The bath wavepacket $\vert G_{j t}^0\rangle$ has a superscript
$0$ to indicate that it is unperturbed from its trajectory if the
system travels in the left arm.  Note that $\vert g_{0t}^l\rangle$
does not interact with the bath and therefore does not develop an
index $i$ depending on the bath state.

When $\hat \rho_{red}$ is squared and traced over to obtain the
decoherence measure ${\cal M}\equiv {\rm Tr}\left [\hat \rho_{\rm
red}^2 \right]$, the terms ${\cal M}_{\rm coh}\equiv {\rm Tr}\left
[\hat \rho_{\rm red}^{rl} \hat \rho_{\rm red}^{lr} +\hat \rho_{\rm
red}^{lr} \hat \rho_{\rm red}^{rl}\right]$ contain all the information
on inter-arm coherence.  These give 
\begin{eqnarray} 
\label{central}
{\cal M}_{\rm coh}& =& {1\over 2} \sum\limits_{ij i' j'} \langle g_{j' t}^r\vert g_{j t}^r\rangle w(i',j') w(j,i) {\cal O}_{j' i' n}^{r 0}{\cal O}_{ij}^{0 r} e^{i\phi_j-i\phi_{j'}} \nonumber \\
& \equiv & \langle \Psi^{\rm sys}\vert \Psi^{\rm sys}\rangle\;,
\end{eqnarray} 
with
\begin{equation} 
\vert \Psi^{\rm sys}\rangle ={1\over \sqrt 2} \sum\limits_{i,j}
w(j,i) {\cal O}_{ij}^{0 r} e^{i\phi_j} \vert g_{jt}^r\rangle\;.
\label{eq:eff_wavefunction}
\end{equation}
Remarkably, ${\cal M}_{\rm coh}$ is the self-overlap of a (generally
non-normalized) effective system wavefunction. The emergence of a
wavefunction form is unexpected because we have not specified that the
bath was initially in a pure state; it may be in a mixed state such as
a thermal bath. We can check Eq.~(\ref{eq:eff_wavefunction}) in the
limit of no interaction with the bath: then, the overlap factors are
all unity, the phases $\phi_j = 0$, and all the gaussians $\vert
g_{jt}^r\rangle$ are the same (normalized) unperturbed state $\vert
g_{0t}^r\rangle$. Then,
\begin{equation} 
\vert \Psi^{\rm sys}\rangle  = {1\over \sqrt 2}\sum\limits_{i}  w(i,i) \vert g_{0t}^r\rangle = 
{1\over \sqrt 2}  \vert g_{0t}^r\rangle\;,
\end{equation} 
implying $ \langle \Psi^{\rm sys}\vert \Psi^{\rm sys}\rangle = {\cal
M}_{\rm coh}= 1/2$, i.e. maximum coherence.  

Eqs.~(\ref{central}) and (\ref{eq:eff_wavefunction}) are the central
results of this section and they are the main formulas of this
paper. They apply to any bath (harmonic or not, pure state or mixed
state) with weak system-bath coupling and any number of total degrees of
freedom.

Eq.~(\ref{central}) states that the system coherence is determined by
an effective system wavefunction with wavepacket overlap factors
$\langle g_{j' t}^r\vert g_{j t}^r\rangle$, bath overlap factors
${\cal O}_{ij}^{0r}$, weighting factors (coming from the initial bath
conditions) $w(j,i)$, and finally phase factors $ e^{i\phi_j}$.  The
phases $\phi_j$ are classical actions divided by $\hbar$ and are given
by Eq.~(\ref{pha}).  The ``system wavefunction'' overlap form,
Eq.~(\ref{eq:eff_wavefunction}), is especially convenient for
intuition and computation.  Decoherence shows up as a reduction in the
norm of the system wavefunction.  This comes about from any or all of
three factors: bath overlap decay, phase jitter, and system wavepacket
displacement.

The are several more interesting facets of Eq.~(\ref{central})
deserving discussion. We will do this systematically, by considering
important special cases which highlight aspects of this formula.

\section{Special Cases of Decoherence}
\label{sec:cases}
We begin our discussion of various limits by assuming that the system
overlaps and possibly also the bath overlaps are near unity. This
regime is indeed accessible, since the classical action perturbation
term is $\phi_i ={1\over \hbar} \delta S_t $, where $\delta S_t$ is
the action due to the perturbation along the unperturbed orbit. There
is no doubt the action term can be large compared to $2 \pi$ and vary
widely, since the perturbing classical action can be large compared to
$\hbar$.  At the same time, the wavepacket displacement can remain small
compared to its width, in both position and momentum space. Classical
action changes are always accompanied by corresponding areas or
volumes in phase space; if one plots the manifolds of the perturbed
system exactly, then a phase $\phi = 2\pi$ will be accompanied by a
loop or area in phase space which is of this magnitude. However, the
wavepacket width goes as $\sim \sqrt\hbar$, but the perturbing action
increases as $\hbar^{-1}$.  Therefore, for small enough perturbations
and small enough $\hbar$, we can safely take the wavepacket overlaps
to be 1, and focus on the phase terms.

Suppose that the system wavepackets $\vert g_{j t}^r\rangle $ are not
displaced by the interaction with the bath, then we have
\begin{equation}
 \langle g_{j' t}^r\vert g_{j t}^r\rangle \approx 1
\label{sys}
\end{equation}
and Eq.~(\ref{central})  becomes 
\begin{eqnarray} 
\label{bathcentral}
\nonumber
{\cal M}_{\rm coh}& =& {1\over 2} \sum\limits_{ij i' j'} w(i',j') w(j,i) {\cal O}_{j' i'}^{r 0}{\cal O}_{ij}^{0 r} e^{i\phi_j-i\phi_{j'}}\\
&=& {1\over 2} \left \vert \sum\limits_{i,j} w(j,i)   {\cal O}_{ij}^{0r} e^{i\phi_j}\right \vert^2\equiv {1\over 2}\vert \mu\vert^2\;.
\end{eqnarray}

If our bath had initially been in a {\it pure state}, $w(j,i)=w_j
w_i^*$ so that
\begin{eqnarray}
\mu&=&\sum_ {ij} w^*_i {\cal O}_{ij}^{0r} e^{i\phi_j} w_j\nonumber \\
&=& \int d\eta \, \chi_l^*(\eta)\chi_r(\eta) \;,
\label{eq:mu}
\end{eqnarray}
which is just a simple bath overlap. Here $\chi_{r (l)}$ is the state
of the bath if the particle went around the right (left) arm; $\eta$
represents the set of bath coordinates.  This is the notation of
Appendix~\ref{app:imry}. 

In many physical situations, it may be the
case that the largest contribution to the sum in Eq.~(\ref{eq:mu}) comes
from $i=j$.  (This might be the case because the bath overlaps ${\cal
O}_{ij}^{0r}$ are small for $i \neq j$ and/or because the off-diagonal terms in
$w^*_iw_j$ oscillate in sign from term to term.)  Making this approximation,
we find
\begin{eqnarray}
\mu& \approx &\sum_ {i} |w_i|^2 e^{i\phi_i} {\cal O}_{ii}^{0r}\;.
\label{eq:mu_approx}
\end{eqnarray}
This gives the interpretation of ${\cal M}_{\rm coh}$ as one half the
modulus squared of a phase factor {\it times a bath overlap} averaged
over different ``runs'', or realizations of the bath, i.e.
\begin{equation} 
{\cal M}_{\rm coh}  = {1\over 2}\vert < e^{i\phi_i} {\cal O}_{ii}^{0r}>_{w_i}\vert ^2\;.
\end{equation}
From the system wavefunction viewpoint, we have (in the
limit of Eq.~(\ref{sys})),
\begin{equation} 
\vert \Psi^{\rm sys} \rangle = \left ( {1\over \sqrt 2}\sum\limits_{i} |w_i|^2 e^{i\phi_i} {\cal O}_{ii}^{0r} \right ) \vert g_{0t}\rangle \;,
\end{equation} 
which lays the blame for decoherence entirely in the sum contained in
the parentheses.  This can be reduced in magnitude by both bath overlap
decay factors or by phase jitter.

\subsection{Nondynamical Bath}
\label{ssec:nondyn}
An important special case to consider is one in which the bath does
not have any dynamics of its own, the so called ``nondynamical'' bath.
The ``nondynamical'' bath limit emerges by further setting ${\cal
O}_{ij}^{0r} = \delta_{ij}$ in Eq.~(\ref{bathcentral}) (or ${\cal
O}_{ij}^{0r} = 1$ in Eq.~(\ref{eq:mu_approx})), i.e. bath wavepackets
undisplaced by the interaction, which would be the case indeed if the
bath Hamiltonian commutes with the bath-particle interaction
potential.  Then,
\begin{equation} 
\vert \Psi^{\rm sys} \rangle = \left ( {1\over \sqrt 2}\sum\limits_{i} |w_i|^2  e^{i\phi_i}\right ) \vert g_{0t}\rangle \;.
\end{equation}
The reduction of the norm (corresponding to decoherence) is due
entirely to phase jitter.  In this case we have a compelling formula
emerging, in the spirit of SAI (See Appendix~\ref{app:imry}),
\begin{equation}
\label{special2}
\int d\eta \, \chi_l^*(\eta)\chi_r(\eta) = \sum\limits_{i} |w_i|^2    e^{i\phi_i}\equiv < e^{i\phi_i} >\;,
\end{equation}
where the phase $\phi_i$ is imparted with probability $|w_i|^2$:
$\phi_i$ is the phase acquired if the bath wavepacket is $\vert
G_{i0}\rangle$ initially, and this happens with probability $|w_i|^2$,
the probability weight of that wavepacket in the initial bath.  This
formula gives a concrete picture of the nondynamical bath limit,
and the origin of the phases which are averaged over: they are
classical actions for the trajectory of the system-bath dynamics,
divided by $\hbar$. In terms of ${\cal M}_{\rm coh}$, we have
\begin{equation}
 {\cal M}_{\rm coh}=\frac{1}{2} \vert <e^{i\phi_i}>\vert^2 \;.
\end{equation}

The limit of a nondynamical bath can also be achieved (without the
bath Hamiltonian commuting with the bath-particle interaction
potential) by a high temperature bath whose wavepacket description
involves mostly very excited coherent states.  Such coherent states
are robust against self overlap decay, unless large energy exchange is
occurring.  This kind of nondynamical bath corresponds to classical
states of the radiation field in a large cavity with high enough
temperature.  It is well known that this situation is described
quantum mechanically in terms of excited coherent states of the field
oscillators.  Such baths (or similarly, externally applied fields)
tend not to contribute to decoherence {\it via} a bath overlap decay
mechanism for weak coupling, but rather the sort of ``dephasing'' 
expressed by Eq.~(\ref{special2}).

\subsection{Dynamical Bath}
\label{ssec:dyn}
More generally, we see from Eq.~(\ref{bathcentral}) that the
decoherence (still assuming little system displacement) arises from
two sources: phase decoherence and amplitude decoherence (due to $\vert
{\cal O}_{ii}^{0r}\vert <1$). The latter is caused by the bath wavepackets
becoming displaced by interaction with the system.  In this case, it
is less compelling to associate $\langle e^{i\phi_i}\rangle $ with $
\sum\limits_{i} |w_i|^2 {\cal O}_{ii}^{0r} e^{i\phi_i}$, since the overlap
factors are not naturally written as integrals over phase factors,
although one could always do this, however absent of physical
motivation. This situation corresponds to SAI's dynamical bath.  The
present formulation shows a pure phase average picture for this case
is somewhat forced. Bath wavepacket displacement (and in the next
section, system wavepacket displacement) thus emerges as a restraint
on a pure ``dephasing'' picture of decoherence.
 
The dynamical bath limit would be uninteresting if decoherence is
always dominated by dephasing.  But low temperature baths are prime
suspects for overlap decay to dominate dephasing effects.  For
example, if there is just one bath coherent state, e.g. as in a $T=0$
bath whose true ground state is describe well by a single
multidimensional coherent state, the Eq.~(\ref{central}) becomes
\begin{equation} 
{\cal M}_{\rm coh} \approx {1\over 2} \vert {\cal O}_{11}^{0r}\vert^2\;,
\end{equation} 
i.e. the decoherence is entirely caused by bath overlap decay.  This
is true even if the system wavepacket is strongly displaced, since the
system wavepacket simply overlaps itself in Eq.~(\ref{central}) when
there is but a single state in the sums.  It is therefore possible to
decohere from the zero temperature initial state or a given single
coherent state of the bath, due to bath overlap decay; however, at zero 
temperature this requires degeneracies of the bath.

Here, it is especially clear that a phase average picture is not
natural for a dynamical bath: in this case there is only one ``quantum
trajectory'' so to speak, a single product wavepacket which describes
the bath-system evolution in the right arm. Eq.~(\ref{central}) shows
that the effect in this case is decoherence due to a displaced bath
wavepacket.

\subsection{System Overlap Decoherence}
\label{ssec:overlap}
We now relax the artificial (though possible) condition that the
system overlap terms are essentially $1$, i.e. the system wavepackets
can be significantly displaced by interacting with the bath. The
concepts of a dynamical and nondynamical bath still apply; the
question at hand now is the further role of system overlap decay in
forming the coherence measure ${\cal M}_{\rm coh}$.  We assume a
nondynamical bath to make the analysis simpler; this means that bath
wavepacket self-overlaps are unity, and any decoherence can be blamed
either on random phases (the dephasing limit) and/or on system overlap
decay. We now investigate the relative importance of these two.

The relevant effective system wavefunction is then
\begin{equation} 
\vert \Psi^{\rm sys}\rangle ={1\over \sqrt 2}  \sum\limits_i |w_i|^2 e^{i \phi_i} \vert g_{it}^r\rangle\;.
\end{equation} 
The self overlap of this wavefunction is 
\begin{equation} 
{\cal M}_{\rm coh} = {1\over 2} \sum\limits_{ij}|w_i|^2 |w_j|^2\ e^{i \phi_i-i\phi_j} \langle g_{jt}^r \vert g_{it}^r\rangle\;.
\end{equation}

The phase and overlap contributions are manifest. It is not possible
to give a general rendition of the relative importance of phase and
overlap contributions to this expression; this will depend on the
system and bath under consideration.  However, the diagonal terms
always survive, even in the limit of strong kicking of the system
wavepackets. (We remind the reader that even though our analysis was
perturbative for the wavepacket displacements, in the sense of
classical perturbation theory, the displacements can be strong in the
quantum sense. ``Strong'' is measured by wavepacket overlap decay,
which can be severe even while classical perturbations are correctly
giving the wavepacket displacements.  See Appendix~\ref{app:gauss}.)
Restoring the bath overlap factors for a moment, for the diagonal
terms we get
\begin{equation}
{\cal M}_{\rm coh} \approx {1\over 2} \sum\limits_i |w_i|^4 \vert {\cal O}_{ii}^{0r} \vert^2 \le {1\over 2} \sum\limits_i |w_i|^4 \;.
\end{equation}
Since  $\sum\limits_i |w_i|^2= 1$, 
\begin{equation}
\sum\limits_i |w_i|^4 \sim 1/N\;,
\label{eq:coherence_estimate}
\end{equation}
is an inverse participation ratio, where $N$ is the number of
participating quantum states describing the bath.  When $N$ is large,
Eq.~(\ref{eq:coherence_estimate}) predicts that ${\cal M}_{\rm coh}$
will be vastly smaller than $1/2$, effectively meaning the system is
completely decohered in this limit. This is the limiting form in the
strong system-kicking limit and makes physical sense: the broader the
distribution of quantum states in the initial bath (measured by $N$ in
Eq.~(\ref{eq:coherence_estimate}), the more uncertain the
``potential'' felt by the system (bath has a broad distribution of
possible states) and hence the greater the decoherence.

There remains a question, however: could the system overlap decay
strongly as above without strong phase randomization, so that a pure
dephasing picture would miss it?  When one considers, for example, the
harmonic model, the conclusion is soon reached that for finite
temperature it is not easy to strongly displace the system wavepackets
randomly without strong phase randomization.

\section{Conclusions}
\label{sec:conclusions}
In this paper, we have presented a semiclassical, wavepacket-based
formalism for decoherence.  We have limited ourselves to the case of a
single system wavepacket split initially into two mutually coherent
pieces, one of which interacts with a bath.  We derive an expression
for the measure of coherence in the system, Eq.~(\ref{central}), which
determines the coherence in terms of wavepacket overlap factors
$\langle g_{j' t}^r\vert g_{j t}^r\rangle$, bath overlap factors
${\cal O}_{ij}^{0r}$, weighting factors (coming from the initial bath
conditions) $w(j,i)$, and finally phase factors $ e^{i\phi_j}$.  The
phases $\phi_j$ are classical actions divided by $\hbar$ and are given
by Eq.~(\ref{pha}).  

One perhaps surprising and potentially very computationally and
intuitively useful aspect of our formulation is the emergence of an
effective system wavefunction, which measures the decoherence,
Eq.~(\ref{eq:eff_wavefunction}): Decoherence shows up as a reduction
in the norm of the system wavefunction. Similar ideas have been
introduced in the context of a stochastic Schrodinger
equation~\cite{strunz}.

After the derivation of the general formulas for the coherence of a
quantum system interacting with a bath in Sec.~\ref{sec:theory}, we
discuss several special limits of the interaction.  In one limit,
discussed in Sec.~\ref{ssec:nondyn}, neither system nor bath
wavepackets are significantly displaced by the interaction, but a
distribution of phases develop which decohere the system.  This is
limit is naturally described as ``dephasing'' and is appropriate to a
number of physical situations where the system-bath interactions are
quite weak.  In Sec.~\ref{ssec:dyn} a situation is discussed where
system wavepackets are barely perturbed but bath wavepackets are
significantly displaced. In this limit, one can still force a random
phase picture, but the identification with an average over random
phase factors is more of a mathematical equivalence than a physically
motivated idea. Finally, if the system wavepacket is strongly
perturbed by the interaction, as in Sec.~\ref{ssec:overlap}, a new
decoherence mechanism sets in: system overlap decay.  Such system
disturbance is hardly rare or unlikely.  Strong perturbation of the
system can occur with or without significant bath displacement.  

Our perturbation treatment has certain similarities to linear response
theory. For chaotic systems (expected for say a liquid or gaseous
bath) it could suffer the same criticism~\cite{vanK} that the actual
magnitude of the perturbation for which the formalism is valid is
unreasonably small.  However, it might benefit from the same saving
graces as linear response theory; namely, that ensembles of
trajectories are better behaved than individual trajectories.

The distinction we are making between phase randomization versus
overlap decay has long been central within the context of spectroscopy
of systems embedded in a bath (see, e.g. Ref. \cite{skinner}).  The
concepts of ``dephasing'', ``depopulation'', and ``pure dephasing''
are traditional in spectroscopy.  Within the context of exponential
decay, the relation
\begin{equation} 
{1\over T_2} = {1\over 2 T_1} + {1\over T_2^*}
\end{equation} 
is legion, where $T_2$ is the dephasing time, $T_1$ is the
depopulation time, and $T_2^*$ is the pure dephasing time.  The
translation of concepts into the present discussion is: ``dephasing''
$\to $ ``decoherence''; ``pure dephasing'' $\to $ ``dephasing''.  The
time $T_2$ is typically the time constant for decay of the initial
wavefunction created by absorption of a photon, and is measured from
the width of of an absorption line. (If we had introduced a tunnel
coupling between the two arms of our model device, we could also have
had a natural population decay-the probability of being in each arm.
This is an interesting subject for future study).

The approach we have taken to decoherence is not limited to the
physical circumstances used here.  The semiclassical
wavepacket-perturbation approach should be applicable to wide variety
of situations and physical measurables including electron decoherence
in metals\cite{chakr,altshuler,zaikin,cohen} and studies of the
classical-quantum correspondence\cite{bertet}.  We hope to pursue some
of these in the near future.

\begin{acknowledgments}
We thank Y. Imry, A. Johnson, D. Reichman, A. Stern, S. Tomsovic,
J. Vanic\'ek and T. Van Voorhis for useful discussions.  This work was
supported by Harvard ITAMP, NSF CHE-0073544, PHY-0117795 and
PHY-9907949.  Finally, EJH would like to thank the Max Planck
Institute for Complex Systems in Dresden for its hospitality, and the
Alexander von Humboldt Foundation for support.
\end{acknowledgments}

\appendix
\section{Coherent States and Gaussian Wavepackets}
\label{app:gauss}
We present a brief review of Gaussian wavepacket dynamics
for our approach to the decoherence problem.  It is well known
that the problem of the usual kinetic energy operator with a time
dependent potential at most quadratic in the coordinates is exactly
solvable, and is especially simple in the case of initial
wavefunctions which are Gaussian; these remain exactly Gaussian
wavepackets under the time evolution.

We note that our goals extend far beyond such
quadratic systems; we will see that a semiclassical approximation
permits the use of quadratic form dynamics in more general contexts.
We make use of the so-called ``thawed guassian
approximation''\cite{thefirst,thesecond,thethird} which employs the
auxiliary variables ${\bf Z}$ and ${\bf P_Z},$ whose dynamics are
given by the equations below.  The ``thawed Gaussian approximation''
allows one to approximate the potential {\em locally} as quadratic
thus taking advantage of the exactness of Gaussian propagation on
quadratic potentials.

In a multidimensional form, a general Gaussian wavepacket is given by
\begin{eqnarray}  
\label{gau}
\psi({\bf q},t)=
\exp \{ {i\over\hbar}[  ( &&{\bf q} -  {\bf q_t})^T\cdot {\bf A_t}
\cdot (  {\bf q} -  {\bf q _t})\nonumber \\
&&
+  {\bf p _t}\cdot (  {\bf q} -  {\bf q_t}) +  s_t]\}\;, 
\label{eq:guassian_wp}
\end{eqnarray} 
where $ {\bf A}_t$ is an $N \times N$ matrix for $N$ coordinates
describing the stability of the center of the Gaussian wavepacket, and
${\bf q_t},\, {\bf p_t}$ are $N$-dimensional vectors describing the
position and momentum evolution of the center of the wavepacket.  We
have introduced the more conventional wavepacket notation ${\bf
q_t}={\bf q}(t)$, etc.  Let the classical Hamiltonian, ${\rm H = T + V
}$, have the usual Cartesian kinetic energy operator and a general
time dependent potential ${\rm V}$ smooth at least up to quadratic
order in the coordinates.  The parameters of the Gaussian then
obey\cite{tdwp,leshouches}
\begin{eqnarray}  
\label{summary}
{d\over dt}\bf q_t&=& \,\nabla_p {\rm H} \\
\label{summary2}
{d\over dt}\bf p_t &=& \, -\nabla_q {\rm H} \\
\label{summary3}
{\bf A_t} &=& {1\over 2}{{\bf P_Z}\cdot {\bf Z^{-1}}} \\
\label{summary4}
{d\over dt}\left(\begin{array}{c}
{\bf P_Z}\\
{\bf Z}
\end{array}\right)
 &=& \, \left( \begin{array}{cc}
{\bf 0}&{-{\bf V''}{(t)}}  \\
{\bf m^{-1}}&{\bf 0} \end{array}
\right )\left(
\begin{array}{c}
{\bf P_Z}\\
 {\bf Z}
\end{array}
\right)\\
\label{summary5}
\dot s_t &=& \,{ L_t} + {i\hbar\over 2} {\rm Tr[ \dot {\bf Z}\cdot{\bf Z^{-1}}]}\;,
\end{eqnarray} 
where $ L_t$ is the usual classical Lagrangian.
Integrating over time, we have
\begin{eqnarray}  
\label{stee}
s_t &=&\, s_0 + S_t + {i\hbar\over 2} {\rm Tr[ \ln{\bf  Z}]}\;,
\end{eqnarray} 
where ${\bf V''}$ and ${\bf m^{-1}}$ are N-dimensional matrices of mixed 
second derivatives 
of the Hamiltonian with respect to position and momentum coordinates, 
respectively.
That is, 
\begin{eqnarray}  
\left[ {\bf V''}\right ]_{ij} = {\partial^2 {  H}\over \partial q_i\partial q_j}\;,
\end{eqnarray} 
and so forth. $S_t$ is the usual classical action.
Eqs.~(\ref{summary}-\ref{summary5}) holds for a general time dependent
${\rm V}$.  We focus on the stability equations, Eq.~(\ref{summary4}),
which admit the solution
\begin{eqnarray} 
\label{pzzeqn}
\left(\begin{array}{c}
{\bf P_{Z_t}}\\
{\bf Z_t}
\end{array}\right) = {\bf M}(t) \left(\begin{array}{c}
{\bf P_{Z_0}}\\
{\bf Z_0}
\end{array}\right)\;,
\end{eqnarray} 
where 
\begin{eqnarray} 
\label{meqn}
{\bf M}(t) = {\hat T}\ e^{\int^t {\bf K}(t') \ dt'}\;,
\end{eqnarray} 
and $ {\hat T}$ denotes the time ordering operator, needed because 
\begin{eqnarray} 
\label{keqn}
 {\bf K}(t) =  \left( \begin{array}{cc}
{\bf 0}&{-{\bf V''}(t)}  \\
{\bf m^{-1}}&{\bf 0} \end{array}
\right )
\end{eqnarray} 
does not commute with itself (in general) at different times.
${\bf M} (t)$ is the usual classical stability matrix, 
where ${\bf M}_{11} = \partial p_t/\partial p_0$, etc.

\begin{figure}
\centering
\includegraphics[width=8cm]{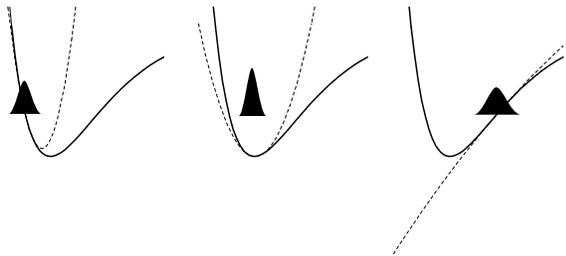}
\caption{A Gaussian wavepacket in an anharmonic potential is
approximately propagated by expanding the potential to quadratic order
locally around the instantaneous center of the Gaussian.}
\label{fig2}
\end{figure}

Consider a narrow (in {\bf q}) Gaussian wavepacket centered on the
classical position ${\bf q_0}$ and momentum ${\bf p_0}$.  Assuming a
reasonably smooth potential, let us expand around ${\bf q_0}$ up to
quadratic terms, arguing that the tails of the Gaussian are negligible
where the Taylor expansion starts to break down.  We use this
quadratic form to propagate the packet in the next time instant. Thus
the Gaussian will propagate in the next instant according to
Eqs.~(\ref{summary}-\ref{summary5}). If we agree to move the center of
the Taylor expansion to the moving mean position of the wavepacket,
${\bf q_t}$, then Eqs.~(\ref{summary}-\ref{summary5}) will hold, since
the potential is now, by construction, a time dependent quadratic form.
However, the interpretation has changed--the position and momentum
parameters ${\bf q_t}$ and ${\bf p_t}$ are now just exactly the usual
classical trajectories on the {\it exact, anharmonic} potential, but
the distortion of the {\em wavepacket is governed by the local quadratic
expansion} of the potential--thus keeping the wavepacket
Gaussian\cite{thefirst}.  We illustrate the idea in Fig.~\ref{fig2}.

In general this approximation breaks down after some time due to
wavepacket spreading, but that time can be put off as long as we
please as $\hbar \to 0$, since we can take a narrower wavepacket, with
position and momentum uncertainties going as $\sqrt \hbar$. This
delays the spreading by at least a factor $\sim 1/ \vert \log
\hbar\vert $ in time (for chaotic
systems)\cite{logtime,logtime2,logtime3,logtime4}.

Since any quantum state (aside from spin states) can be built out of
Gaussians, we have a full semiclassical approach, exact as $\hbar \to
0$.  Each Gaussian is propagated with its own optimized time-dependent
Hamiltonian.

The phase $s_t$ of Eq.~(\ref{stee}) is the usual action, taken along
the guiding trajectory, modified by an extra term which takes the
place of a Maslov phase.  This term evolves smoothly in time and therefore is
another advantage of a wavepacket approach as compared to the more
troublesome eigenfunctions of Hermitian operators.

There are other approaches based on wavepackets which share the
smoothing property, such as a full semiclassical coherent state
propagator\cite{weissman1,weissman2,cohst:little}.  This differs from
the above Gaussian wavepacket (GWP) method in that the propagator is
optimized for both the initial Gaussian and a final Gaussian onto
which it is projected.  It is also the ``natural'' semiclassical
approximation if one retains stationary phase as the defining idea
while passing to a coherent state basis. The coherent state method is
usually more accurate at the same value of $\hbar$ compared to GWP,
but it suffers several complexities.  Foremost among these is that
complex classical trajectories come in, (albeit for real time) with
their attendant analytical difficulties including Stokes
lines. Therefore, we choose the GWP approach as a best compromise
between accuracy and complexity.  In the semiclassical limit, $\hbar
\to 0$, there is no accuracy compromise.

Perturbation theory may seem limited in scope at first glance, but we
soon realize that strong interaction with many bath degrees of freedom
leads to such rapid and near complete decoherence that an analysis of the
strong system-bath coupling regime seems a little {\it post
mortem}. We therefore focus our attention on the weak coupling limit.

In the setup in Fig.~(\ref{fig1}), we have a Hamiltonian given by 
\begin{equation}
\hat H=\hat H_{\rm p} + \hat H_{\rm b} + \hat V \equiv \hat H_0 +\lambda \hat V_1 \;,
\label{eq:hamiltonian1}
\end{equation}
where $\hat H_{\rm p}$ ($\hat H_{\rm b}$) is the Hamiltonian of the
particle (bath) and $\hat V$ is the coupling between them.  Suppose we
have solved the $\hat H_0=\hat H_{\rm p} + \hat H_{\rm b}$ problem,
and now we wish to include the effects of $\lambda \hat V_1$, the
system-bath interaction. In accordance with perturbation theory, the
first order effect is determined by the extra potential felt by the
wavepacket as it travels over its old trajectory. Assuming this
perturbation is smooth as a function of coordinates, we include only
terms linear in the interaction strength, $\lambda$. We can include
perturbations to quadratic order as well, at the cost of increased
complexity, but we defer this for future work.  A weak interaction
with a bath will show up in two ways in the wavepacket given in
Eq.~(\ref{eq:guassian_wp}): (1) Changes to the guiding trajectory,
{\bf $q_t$} and {\bf $p_t$} and (2) changes to the phase $s_t$.

\subsubsection{Perturbation of the Guiding Trajectory}

Let ${\bf q_0}(t)$ be the solution of the $\hat H_0$ problem for a
particular trajectory. The first order perturbed solution we take to
be ${\bf q}(t) = {\bf q_0}(t) + \lambda \delta {\bf q}(t)$, ${\bf
p}(t) = {\bf p_0}(t) + \lambda \delta {\bf p}(t)$. By substituting
this into Hamilton's equations, we find that $\delta {\bf q_t}, \delta
{\bf p_t} $ obey
\begin{eqnarray} 
\label{lin2}
{d\over dt}\left(\begin{array}{c}
{\bf \delta p_t}\\
{\bf \delta q_t}
\end{array}\right)
 &=& \, \left( \begin{array}{cc}
{\bf 0}&{-{\bf V''}(t)}  \\
{\bf m^{-1}}&{\bf 0} \end{array}
\right )\left(
\begin{array}{c}
{\bf \delta p_t}\\
 {\bf \delta q_t}
\end{array} \right)+ \left (\begin{array}{c}
{\bf f}(t)\\
 {\bf 0} 
\end{array}\right)\;, \nonumber \\
\end{eqnarray} 
where $f(t) = - V_1^\prime({\bf q_0}(t))$ is the time dependent
forcing function. The solution is
\begin{eqnarray} 
\label{lin3}
\left(\begin{array}{c}
{\bf \delta p_t}\\
{\bf \delta q_t}
\end{array}\right)
& =& 
\label{lin4}
{\bf \large M}(t) \left(
\begin{array}{c}
{\bf \delta p_0}\\
 {\bf \delta q_0}
\end{array} \right)
+  {\bf \large M}(t) \int\limits_0^t {\bf M}(t')^{-1}\left (\begin{array}{c}
{\bf f}(t')\\
 {\bf 0} 
\end{array}\right) dt'\nonumber \\ \\
&=& {\bf \large M}(t) \int\limits_0^t {\bf M}(t')^{-1}\left (\begin{array}{c}
{\bf f}(t')\\
 {\bf 0} 
\end{array}\right)\ dt', 
\end{eqnarray} 
since $\bf \delta p_0 = \delta q_0 = 0$ 
in the present circumstances, with
\begin{eqnarray} 
{\bf M}(t)^{-1} = {\hat T}^\dagger\ e^{-\int^t {\bf K}(t') \ dt'}\;,
\end{eqnarray} 
and $ {\hat T}^\dagger$ forces the reverse order of times;
i.e. earliest times to the left in the series expansion of ${\bf
M}(t)^{-1} $. Note that $(d/dt) {\bf M}(t)^{-1} = {\bf M}(t)^{-1} {\bf
K}(t)$.  Thus, the perturbation of a general trajectory in classical
mechanics generates a linearly forced oscillator problem. This is not
surprising, since the small additional potential creates small new
forces on the particle near its old trajectory.

\subsubsection{Perturbation of the Phase}

To see how the phase of the wavepacket changes under the perturbation, we 
examine 
\begin{eqnarray} 
\delta s_t &=& \delta S_t \nonumber \\
&=& \delta \int (T-V_0 - \lambda  V_1)\ dt'\nonumber \\  
&=& \delta \int (T-V_0)\ dt'  - \delta \int \lambda V_1\ dt' \;.
\end{eqnarray} 
We see the change comes in two parts.  The first is the change in the ``old'' action (it involves the old $V_0$) due to the new trajectory.  Assuming the new one has not wandered too far, we have 
\begin{eqnarray}
\delta \int (T-V_0) dt' = \lambda  \ p_t \delta q_t\;,
\end{eqnarray} 
by the stationary principle of the action (the RHS of this equation is not zero since 
the perturbation causes a drift in final position, $ \lambda  \delta q_t$.)
Thus, the   wavepacket phase change, to first order, is 
\begin{eqnarray} 
\label{pha}
\phi_i &=& \frac{\delta s_t}{\hbar}\nonumber \\
&=&\frac{1}{\hbar} \delta \int (T-V_0 - \lambda  V_1)\ dt' \nonumber \\
& =&\lambda  \  \frac{p_t \delta q_t}{\hbar}\ - \frac{\lambda}{\hbar}  \int  V_1(q_0(t'))\ dt'\;.
\end{eqnarray}

\section{The Dephasing Arguments of Stern, Aharonov and Imry}
\label{app:imry}
Stern, Aharonov and Imry (SAI) have argued\cite{ady,ady:nato} that
decoherence may be thought of as dephasing, i.e.  that a quantum
particle acquires a broad distribution of possible phases so that its
phase is ``randomized'' and thereby loses coherence.  SAI argue that
decoherence due to shifting a bath into an orthogonal state and
decoherence due to a particle acquiring a broad distribution of
possible phases are equivalent.  In this appendix we give the skeleton
of their argument in their original notation to provide a counter
point for our discussion in the body of this paper.

SAI consider a quantum particle (coordinate $x$) moving around both
arms of an Aharonov-Bohm ring threaded by magnetic flux with a bath
(coordinate $\eta$) that only interacts with the particle in the right
arm as shown in Fig.~\ref{fig1}. Particles moving around the left arm
are assumed not to interact with the environment.  The initial
wavefunction is taken to be
\begin{equation}
\psi(t=0)=[l(x)+r(x)]\otimes\chi_0(\eta)\;,
\label{eq:psi_init}
\end{equation}
and corresponds to the particle having just entered the ring region
(near point A in Fig.~\ref{fig1}), but not yet interacting with the
bath. Here $l(x)$ ($r(x)$) is the initial particle wavefunction on the
left (right) arm, assumed to be a wavepacket, and $\chi_0(\eta)$ is
the initial state of the bath, assumed to be localized in the right
arm.  SAI then take a final wavefunction (near point B in
Fig.~\ref{fig1})
\begin{equation}
\psi(\tau)=l(x,\tau)\otimes\chi_l(\eta)+r(x,\tau)\otimes\chi_r(\eta)\;,
\label{eq:main_approx}
\end{equation}
where $\chi_l(\eta)$ ($\chi_r(\eta)$) is the state of the bath had the
particle gone around the left (right) arm\cite{comment_1}.  SAI
distinguish between ``dynamical'' and ``nondynamical'' environments,
which either do or do not involve non-trivial dynamics on their own,
respectively.  From Eq.~(\ref{eq:main_approx}), SAI obtain the result
that the interference term is (taking the trace over the environment)
\begin{equation}
2{\rm Re}\left [l^*(x)r(x)\int d\eta \chi_l^*(\eta)\chi_r(\eta)\right ]\;,
\label{eq:interference}
\end{equation}
which allows one to interpret the reduction of the interference (loss
of coherence) in terms of a reduction in the overlap of the bath
states for the two paths around the ring\cite{comment_imry}. SAI then
argue that one can make the identification
\begin{equation}
\langle e^{i\hat \phi}\rangle=\int d\eta \, \chi_l^*(\eta)\chi_r(\eta)\;,
\label{eq:equiv}
\end{equation}
where $\langle e^{i\hat \phi}\rangle \equiv \langle \chi_0 |e^{i\hat
\phi} |\chi_0 \rangle$ and where for nondynamical environments the
phase angle $\phi$ is
\begin{equation}
\phi=-\int dt\, V[x(t)]/\hbar\;,
\label{eq:phi}
\end{equation}
with $x(t)$ the classical trajectory of the particle around the right 
ring arm\cite{comment_2}. For the case of dynamical environments
\begin{equation}
\langle \chi_0 |e^{i\hat \phi}| \chi_0 \rangle=\langle \chi_0|\hat T e^{-{i \over \hbar} \int_0^{\tau}dt'\,\hat V_I(t')}|\chi_0\rangle \;,
\label{eq:time_order}
\end{equation}
where $\hat V_I(t')$=$e^{i\hat H_{\rm b}t'/\hbar} \hat V e^{-i\hat
H_{\rm b}t'/\hbar}$ is the potential in the interaction representation
and $\hat T$ is the time ordering operator. A nondynamical environment
is distinguished from a dynamical one in that the interaction picture
operator $\hat V_I(t')$ commutes with itself at different times in the
nondynamical case.  For a dynamical environment it is not generally
possible\cite{ady,loss} to write down a simple relationship such as
that expressed in Eq.~(\ref{eq:phi}).  However, for a nondynamical
environment,
\begin{equation}
\langle e^{i\hat \phi}\rangle=\int d\phi P(\phi) e^{i \phi}\;,
\label{eq:phase_dist}
\end{equation}
where $P(\phi)\equiv |\chi_0(\eta(\phi))|^2 {d \eta \over d \phi}$.

The central result of the work of SAI is the equivalence expressed in
Eq.~(\ref{eq:equiv}) which states that the reduction of
Eq.~(\ref{eq:interference}) can be viewed as either the shifting of
the bath states for the two paths around the ring (due to the
interactions in the right arm of the ring) or as the particle on the
interacting arm being subject to an uncertain potential resulting in
an uncertain phase shift and hence a reduction in $\langle e^{i\hat
\phi}\rangle$.

The key approximation of SAI is the assumption that the interaction in
the right arm is weak enough to neglect changes in the trajectory of
the particle, while still allowing the bath to change in response to
the presence of the particle.  This approximation eliminates
entangling on the upper arm (the overall wavefunction is, of course,
still entangled) and results in the appearance of a direct product of
bath states in the interference term, Eq.~(\ref{eq:interference}).
This simplification is necessary to reach a ``pure dephasing"
expression.  In the general situation, the trajectory of the particle
is in fact altered in the interacting arm, which leads to entangling
with the bath, a dephasing effect not included in the SAI picture.

\bibliographystyle{apsrev}
\bibliography{fiete_decoherence.bib}

\begin{thebibliography}{10}
\bibitem{comment_1}
\bibinfo{note}{This innocent looking final wavefunction actually neglects
  entangling on the right arm between the particle and the bath by writing the
  term $r(x,\tau_0)\otimes\chi_r(\eta)$ as a direct product which is not
  correct in general.}
\bibitem{comment_imry}
\bibinfo{note}{One must also also be careful when interpreting the extrema of
  the interference term of Eq.~(\ref{eq:interference}) (as a function of
  magnetic field in the Ahronov-Bohm experiment, for example) as a measure of
  the coherence. The difference of the maximum and minimum of the interference
  term must be normalized by the direct terms to divide out Debye-Waller
  factors which reduce both the the direct terms above and the interference
  cross terms\cite{imry:dephase}. In general one must normalize the
  interference term by a flux dependent term to prevent ``flux blocking"
  effects from being interpreted as decoherence.}
\end{thebibliography}

\end{document}